\documentclass[aps,prd,pdflatex,reprint,10pt,superscriptaddress,floatfix,nofootinbib]{revtex4-2}

\usepackage{float}
\usepackage{amsmath}
\usepackage[T1]{fontenc}
\usepackage[utf8]{inputenc}
\usepackage{lmodern}
\usepackage{amssymb}
\usepackage{natbib}
\usepackage{graphicx}
\usepackage{hyperref}
\usepackage{epstopdf}
\usepackage{pifont}
\usepackage{textgreek}
\usepackage[normalem]{ulem}
\usepackage{placeins}
\usepackage{tikz}
\usepackage{physics}
\usepackage{orcidlink}

\hypersetup{
    colorlinks=true,
    linkcolor=blue,
    citecolor=blue,
    urlcolor=blue
}

\newcommand{\Mren}{\mathcal M}

\begin{document}

\title{Geometry-induced Casimir response in a helicoidal spacetime}

\author{Edilberto O. Silva\orcidlink{0000-0002-0297-5747}}
\email[Edilberto O. Silva - ]{edilberto.silva@ufma.br}
\affiliation{
Programa de P\'os-Gradua\c c\~ao em F\'{\i}sica \&
Coordena\c c\~ao do Curso de F\'{\i}sica -- Bacharelado,
Universidade Federal do Maranh\~{a}o,
65085-580 S\~{a}o Lu\'{\i}s, Maranh\~{a}o, Brazil}

\begin{abstract}
We investigate the Casimir response of a massive scalar field in a torsionless helicoidal spacetime with Levi-Civita connection. The background is ultrastatic and curved, with scalar curvature \(R=-2\Omega^2\), and its off-diagonal metric component induces a geometric coupling between angular and axial quantum numbers. We show that individual modes exhibit a linear helicoidal splitting, whereas the linear contribution cancels in the nonchiral vacuum sum. The leading Casimir correction is therefore quadratic in the twist and defines a helicoidal vacuum susceptibility after local ultraviolet subtractions. For a cylindrical Dirichlet cavity, we compute this scheme-defined susceptibility and the associated correction to the radial Casimir force. The results identify torsionless helicoidal geometry as a controlled setting in which mode-level chirality produces a finite quadratic response of vacuum fluctuations.
\end{abstract}

\maketitle

\section{Introduction}

The Casimir effect is one of the most direct manifestations of the dependence of quantum vacuum fluctuations on boundary conditions, topology, and geometry~\cite{Casimir1948,CasimirPolder1948,Milton2001,Bordag2009,Elizalde1994}. In curved or topologically nontrivial backgrounds, the vacuum spectrum is modified not only by material boundaries but also by the global and local structure of spacetime. Minimal geometries in which these effects can be separated analytically are especially useful: they make it possible to identify which part of the vacuum response is caused by topology, curvature, boundary conditions, or mode mixing.

Cylindrical and conical geometries have played an important role in this context. In cosmic-string backgrounds, the planar angle deficit changes the angular spectrum and induces vacuum polarization even in the absence of material boundaries~\cite{VilenkinShellard1994,HelliwellKonkowski1986,Hiscock1987,Linet1987,FrolovSerebriany1987,Dowker1987,Bordag1990,CognolaKirstenVanzo1994,KhusnutdinovBordag1999}. When coaxial cylindrical boundaries are introduced, the vacuum expectation values may be separated into boundary-free and boundary-induced contributions. This strategy has been applied to scalar, electromagnetic, and fermionic fields in conical and cosmic-string spacetimes~\cite{BezerraDeMello2006PRD,BezerraDeMello2007PLB,BezerraDeMello2008PRD,BezerraDeMello2012PRD}.

Related studies include Casimir effects in compactified cosmic-string backgrounds, parallel-plate configurations in conical spacetimes, Casimir--Polder interactions in cylindrical geometries, and scalar fields in cosmic dispiration spacetimes~\cite{BezerraDeMelloSaharian2012CQG,BezerraDeMelloSaharianGrigoryan2012JPA,BezerraDeMelloBezerraMotaSaharian2012PRD,BezerraEtAl2011EPJC,BezerraDeMelloSaharianAbajyan2018PRD,MotaBezerraDeMelloBakke2018IJMPD}. These works show that topological and geometric parameters can leave direct signatures in vacuum energies, stresses, and forces. They also provide the natural reference point for the present work: a cylindrical Casimir problem in which the background geometry is not conical, but helicoidally twisted.

Helicoidal and screw-dislocation-inspired geometries have also appeared in several quantum-mechanical settings. Early studies of constrained particles on helical strips, helicoidal ribbons, and twisted quantum wires showed that geometric twist can generate effective potentials, chiral transport, and geometry-induced spectral shifts~\cite{DandoloffTruong2004,EntinMagarill2002,MagarillEntin2003JETP,AtanasovDandoloffSaxena2009}. More directly, the spatial part of the metric used below is the left-invariant metric of the three-dimensional Heisenberg group and is closely related to the elastic-Landau-level geometry generated by a uniform density of screw dislocations~\cite{SilvaNettoFurtado2008,BakkeMoraes2012PLA}. Recent applications of this metric and of closely related torsional/dislocation backgrounds include helicoidal quantum wells with geometry-induced chiral currents, torsion-engineered quantum dots and nanowires with nonlinear optical responses, and magnetoelastic Landau quantization~\cite{Silva2026OQE,Silva2026AnnPhys,PereiraSilva2026PhysicaE,AssafraoAhmedSilva2026PhysicaB}. These works establish the spectral and optical relevance of the angular--axial geometric coupling. The present paper asks the complementary quantum-field-theoretic question: how does the same helicoidal mixing affect vacuum fluctuations and the Casimir force of a cylindrical cavity?

In this work we consider the Casimir response in the helicoidal spacetime \cite{SilvaNettoFurtado2008,BakkeMoraes2012PLA,Silva2026AnnPhys,Silva2026OQE}
\begin{equation}
 ds^2=-dt^2+dr^2+r^2d\phi^2+
 \left(dz+\Omega r^2d\phi\right)^2 .
 \label{metric}
\end{equation}
The parameter $\Omega$ measures a metric twist. It is not a Cartan torsion parameter: throughout the paper the connection is Levi-Civita and the Riemann--Cartan torsion tensor is identically zero. Nevertheless, the spacetime is curved, with constant scalar curvature 
\begin{equation}
 R=-2\Omega^2 .
 \label{scalar-curvature}
\end{equation}
Thus the model is neither a flat-space reparametrization nor a torsional defect model. It is a torsionless curved background with an off-diagonal metric component that continuously mixes angular and axial displacements. In contrast with Riemann--Cartan realizations of screw-dislocation media, where torsion is a material defect field, here the twist is treated as a metric property of a Levi-Civita geometry.

This geometry is important for the Casimir problem for three reasons. First, it is ultrastatic and admits the Killing fields $\partial_t$, $\partial_\phi$, and $\partial_z$, so that the scalar wave equation is separable and a natural positive-frequency decomposition exists. Second, the twist preserves cylindrical symmetry while coupling two quantum numbers, the azimuthal number $m$ and the axial momentum $k$. Third, the same parameter that controls this angular--axial coupling also changes the local curvature. The model therefore provides a compact and controlled laboratory for distinguishing curvature effects, radial confinement effects, and chiral mode splitting within a single self-adjoint spectral problem.

The physical question addressed here is how the helicoidal twist modifies the Casimir response and the radial Casimir force of a scalar field confined in a cylindrical cavity. We take the field in the region $0\le r\le R_0$, with real self-adjoint boundary conditions at $r=R_0$. The boundary is cylindrical; the helicoidal character belongs to the spacetime geometry itself. This distinction is essential: the construction does not require a helicoidal material surface, but rather studies how a cylindrical cavity responds when the ambient geometry contains a torsionless helicoidal twist.

The main quantity of interest is the renormalized twist-induced contribution
\begin{equation}
 \Delta E_{\rm Cas}(\Omega)=E_{\rm Cas}(\Omega)-E_{\rm Cas}(0),
\end{equation}
together with the associated radial force,
\begin{equation}
 F_{\rm Cas}=-\frac{\partial E_{\rm Cas}}{\partial R_0}.
\end{equation}
Although individual modes are shifted linearly by the angular--axial coupling, this contribution cancels in the total vacuum energy when the spectrum is summed symmetrically over $m$, $-m$, $k$, and $-k$. Thus, for nonchiral boundary conditions, the leading robust response is quadratic in $\Omega$. This separation between a linear mode-level chirality and a quadratic vacuum-level response is the central mechanism studied below.

The novelty of the present work is not the introduction of another cylindrical Casimir configuration, but the identification of a torsionless helicoidal metric twist as a geometric mechanism that splits individual vacuum modes linearly while producing only an even, quadratic response in the nonchiral Casimir energy. This separates mode-level chirality from vacuum-level symmetry restoration and defines a helicoidal Casimir susceptibility after the standard local ultraviolet subtractions. In this form, the problem fits naturally within quantum field theory in curved spacetime: the background is a curved Levi-Civita geometry, the boundary is a simple self-adjoint cylindrical confinement, and the observable response is a renormalized vacuum-energy correction.

The paper is organized as follows. Section~\ref{sec:spectrum} derives the scalar field equation and formulates the cylindrical spectral problem. Section~\ref{sec:casimir} introduces the zeta-regularized Casimir energy, the small-twist response, and the ultraviolet local-subtraction structure. Section~\ref{sec:numerical} presents Dirichlet spectral benchmarks, finite symmetric mode-sum diagnostics, the extraction of the finite helicoidal susceptibility, and robustness checks. Section~\ref{sec:discussion} discusses the radial force, the radius dependence of the susceptibility, and the physical interpretation, and Sec.~\ref{sec:conclusion} contains the conclusions.

\section{Field equation and cylindrical spectrum}
\label{sec:spectrum}

For the metric \eqref{metric}, the determinant and the inverse components entering the Klein-Gordon operator are
\begin{align}
 &\sqrt{-g}=r,\qquad g^{tt}=-1,\qquad g^{rr}=1,
 \nonumber\\
 &g^{\phi\phi}=\frac1{r^2},\qquad
 g^{\phi z}=-\Omega,\qquad
 g^{zz}=1+\Omega^2r^2 .
 \label{inverse-metric}
\end{align}
A direct computation with the Levi-Civita connection gives the scalar curvature in Eq.~\eqref{scalar-curvature}.

We consider a massive scalar field with nonminimal coupling,
\begin{equation}
 \left(\Box-\mu^2-\xi R\right)\Phi=0,
 \label{KG}
\end{equation}
and separate variables as
\begin{equation}
 \Phi=e^{-i\omega t}e^{im\phi}e^{ikz}F(r),
 \qquad m\in\mathbb Z .
 \label{ansatz}
\end{equation}
If the axial direction is compactified, $z\sim z+L_z$, then $k=2\pi n_z/L_z$. In the numerical part we work with a finite periodic axial length \(L_z\), so that the mode sum is discrete. In the noncompact limit the corresponding quantities should be interpreted per unit axial length, with the sum over \(k\) replaced by the standard continuum integral. Substitution in Eq.~\eqref{KG} gives
\begin{equation}
 F''+\frac1rF'-\frac{m^2}{r^2}F-\Omega^2k^2r^2F+
 \Lambda F=0,
 \label{radial-equation}
\end{equation}
with
\begin{equation}
 \Lambda=\omega^2-\mu^2-\xi R-k^2+2\Omega m k .
 \label{Lambda}
\end{equation}
It is useful to write this equation as the radial eigenvalue problem
\begin{equation}
 \mathcal L_{mk}F=\lambda F,
 \qquad \lambda=\Lambda,
 \label{radial-operator}
\end{equation}
where
\begin{equation}
 \mathcal L_{mk}=-\frac1r\frac{d}{dr}\left(r\frac{d}{dr}\right)
 +\frac{m^2}{r^2}+\Omega^2k^2r^2 .
 \label{radial-L}
\end{equation}
The normal-mode frequencies are therefore
\begin{equation}
 \omega^2=\mu^2-2\xi\Omega^2+k^2-2\Omega m k+\lambda .
 \label{frequency-general}
\end{equation}
This expression displays the three effects of the helicoidal twist: the curvature shift, the radial deformation proportional to $\Omega^2k^2r^2$, and the angular--axial mixing proportional to $\Omega m k$.

\medskip
\noindent\emph{Cylindrical spectral problem.}---The field is confined to the cylindrical region $0\le r\le R_0$. The radial problem is defined in the Hilbert space with measure $r\,dr$,
\begin{equation}
 \langle F_1,F_2\rangle=
 \int_0^{R_0}r\,dr\,F_1^*(r)F_2(r),
\end{equation}
with regularity at the origin and a real boundary condition at $r=R_0$. We use the Robin family
\begin{equation}
 F'(R_0)+\eta F(R_0)=0,
 \qquad \eta\in\mathbb R,
 \label{robin}
\end{equation}
while the Dirichlet case is treated separately as $F(R_0)=0$. These conditions lead to a real self-adjoint spectrum.

For $\Omega k\neq0$, the solution regular at the origin is
\begin{equation}
 F(r)=r^{|m|}e^{-\rho/2}M(a,b,\rho),
 \qquad \rho=|\Omega k|r^2,
 \label{regular-solution}
\end{equation}
where
\begin{equation}
 b=|m|+1,
 \qquad
 a=\frac{|m|+1}{2}-\frac{\lambda}{4|\Omega k|}.
 \label{a-def}
\end{equation}
The Robin spectrum follows from
\begin{equation}
 \mathcal D_m^{(\eta)}(\lambda;\Omega,k,R_0)=0,
 \label{spectral-determinant}
\end{equation}
with
\begin{align}
 \mathcal D_m^{(\eta)}
 &=\left[\frac{d}{dr}+\eta\right]
 \left\{r^{|m|}e^{-|\Omega k|r^2/2}
 M\left(a,b,|\Omega k|r^2\right)\right\}_{r=R_0} .
\end{align}
For Dirichlet boundary conditions the quantization condition simplifies to
\begin{equation}
 M\left(a,b,|\Omega k|R_0^2\right)=0 .
\end{equation}

The untwisted reference spectrum is obtained from the Bessel limit. For $\Omega=0$, or $k=0$,
\begin{equation}
 F(r)=J_{|m|}(\gamma r),
 \qquad \lambda=\gamma^2,
\end{equation}
and Robin boundary conditions give
\begin{equation}
 \gamma J'_{|m|}(\gamma R_0)+\eta J_{|m|}(\gamma R_0)=0.
 \label{bessel-robin}
\end{equation}
This spectrum provides the reference point for the twist-induced Casimir energy difference.

\section{Casimir energy and small-twist response}
\label{sec:casimir}

Let $\lambda_{qmk}(\Omega)$ denote the radial eigenvalues, where $q$ labels the radial excitation. The normal-mode frequencies are
\begin{equation}
 \omega_{qmk}^2(\Omega)=
 \mu^2-2\xi\Omega^2+k^2-2\Omega m k+\lambda_{qmk}(\Omega).
 \label{omega-spectrum}
\end{equation}
The formal zero-point energy,
\begin{equation}
 E_0(\Omega)=\frac12\sum_{q,m,k}\omega_{qmk}(\Omega),
\end{equation}
is divergent. We regularize it through the spectral zeta function
\begin{equation}
 \zeta_\Omega(s)=
 \sum_{q,m,k}\left[\omega_{qmk}^2(\Omega)\right]^{-s},
 \label{zeta-def}
\end{equation}
and define
\begin{equation}
 E(s;\Omega)=\frac12\Mren^{2s}\zeta_\Omega\left(s-\frac12\right),
 \label{regularized-energy}
\end{equation}
where $\Mren$ is a renormalization scale.

The twist-induced contribution is obtained from
\begin{equation}
 \Delta\zeta(s;\Omega)=\zeta_\Omega(s)-\zeta_0(s),
\end{equation}
together with the subtraction of local heat-kernel terms. Since varying $\Omega$ changes both the curvature and the boundary invariants, this renormalization step is essential. In what follows, $\Delta E_{\rm Cas}(\Omega)$ denotes the finite renormalized part.

A contour representation follows from the argument principle:
\begin{align}
 \zeta_\Omega(s)
 &=\sum_{m,k}\frac1{2\pi i}\int_{\mathcal C}d\lambda\,
 \left[\mu^2-2\xi\Omega^2+k^2-2\Omega m k+\lambda\right]^{-s}
 \nonumber\\
 &\quad\times
 \frac{\partial}{\partial\lambda}
 \ln \mathcal D_m^{(\eta)}(\lambda;\Omega,k,R_0),
 \label{zeta-contour}
\end{align}
where $\mathcal C$ encloses the positive zeros of the radial determinant. This representation is suitable for analytic continuation and for numerical evaluation.

\medskip
\noindent\emph{Small-twist expansion.}---The role of the helicoidal parameter is especially transparent for $|\Omega R_0|\ll1$. Let $\gamma_{qm}^2$ be the eigenvalues of the untwisted problem \eqref{bessel-robin}. The unperturbed frequencies are
\begin{equation}
 \omega_{0,qmk}^2=\mu^2+k^2+\gamma_{qm}^2 .
 \label{omega0}
\end{equation}
The radial operator can be written as
\begin{equation}
 \mathcal L_{mk}=\mathcal L_m^{(0)}+\Omega^2k^2r^2,
\end{equation}
where
\begin{equation}
 \mathcal L_m^{(0)}=-\frac1r\frac{d}{dr}\left(r\frac{d}{dr}\right)+\frac{m^2}{r^2}.
\end{equation}
First-order perturbation theory in the operator gives
\begin{equation}
 \lambda_{qmk}(\Omega)=\gamma_{qm}^2+
 \Omega^2k^2\langle r^2\rangle_{qm}+O(\Omega^4),
 \label{lambda-expansion}
\end{equation}
with
\begin{equation}
 \langle r^2\rangle_{qm}=\frac{
 \int_0^{R_0}r\,dr\,r^2|F_{qm}^{(0)}(r)|^2}{
 \int_0^{R_0}r\,dr\,|F_{qm}^{(0)}(r)|^2}.
\end{equation}
Therefore,
\begin{equation}
 \omega_{qmk}^2(\Omega)=\omega_{0,qmk}^2
 -2\Omega m k+
 \Omega^2\left[k^2\langle r^2\rangle_{qm}-2\xi\right]
 +O(\Omega^4).
 \label{omega-small}
\end{equation}
Expanding the square root,
\begin{align}
 \omega_{qmk}(\Omega)
 &=\omega_{0,qmk}
 -\Omega\frac{mk}{\omega_{0,qmk}}
 +\frac{\Omega^2}{2\omega_{0,qmk}}
 \left[k^2\langle r^2\rangle_{qm}-2\xi\right]
 \nonumber\\
 &\quad
 -\frac{\Omega^2m^2k^2}{2\omega_{0,qmk}^3}
 +O(\Omega^3).
 \label{omega-expanded}
\end{align}
For real nonchiral boundary conditions and symmetric sums over $m$ and $k$, the term linear in $\Omega$ vanishes. Hence
\begin{equation}
 \Delta E_{\rm Cas}(\Omega)=\Omega^2\chi_{\rm Cas}+O(\Omega^4).
 \label{DeltaE-quadratic}
\end{equation}
The formal susceptibility is
\begin{equation}
 \chi_{\rm Cas}=\frac14\sum_{q,m,k}^{\rm ren}
 \left[
 \frac{k^2\langle r^2\rangle_{qm}-2\xi}{\omega_{0,qmk}}
 -\frac{m^2k^2}{\omega_{0,qmk}^3}
 \right].
 \label{chi}
\end{equation}
This expression separates the curvature contribution, the radial oscillator-like deformation, and the second-order effect of angular--axial mode mixing.

It is useful to make the scaling explicit. In units with $\hbar=c=1$, the twist has dimension $[\Omega]=L^{-1}$, while $E_{\rm Cas}$ has dimension $L^{-1}$. Therefore the susceptibility in
\begin{equation}
 \Delta E_{\rm Cas}(\Omega)=\Omega^2\chi_{\rm Cas}+O(\Omega^4)
\end{equation}
has dimension of length. For the Dirichlet geometry considered below one may write
\begin{equation}
 \chi_{\rm fin}(R_0,L_z,\mu,\xi)
 =R_0\,\mathcal C(\mu R_0,L_z/R_0,\xi),
 \label{chi-scale}
\end{equation}
where $\mathcal C$ is dimensionless. The numerical values quoted in Sec.~\ref{sec:numerical} correspond to the dimensionless choice $R_0=1$, $L_z=2\pi$, and $\mu=1$. For numerical purposes we also introduce a smooth cutoff version,
\begin{equation}
 \chi_{\epsilon}=\frac14\sum_{q,m,k}
 \left[
 \frac{k^2\langle r^2\rangle_{qm}-2\xi}{\omega_{0,qmk}}
 -\frac{m^2k^2}{\omega_{0,qmk}^3}
 \right]e^{-\epsilon\omega_{0,qmk}} .
 \label{chi-cutoff}
\end{equation}
The limit $\epsilon\to0$ contains local ultraviolet contributions associated with the bulk and the cylindrical boundary. This follows from the standard heat-kernel structure of cutoff-regularized zero-point sums~\cite{Gilkey1995,Kirsten2001,Vassilevich2003}: in three spatial dimensions, the high-frequency part is built from local volume and surface invariants and appears as inverse powers of the cutoff, with possible logarithmic contributions tied to local counterterms and to the renormalization scale. In the present small-twist problem, those local terms are generated at order $\Omega^2$ by the curvature shift, the radial perturbation, and the boundary data.

Equation~\eqref{chi-cutoff} is therefore first used as a controlled cutoff diagnostic. To extract a finite part, we fix a minimal local power-subtraction prescription and model the short-cutoff behavior by
\begin{equation}
 \chi_{\epsilon}=
 \frac{A_4}{\epsilon^4}+\frac{A_3}{\epsilon^3}
 +\frac{A_2}{\epsilon^2}+\frac{A_1}{\epsilon}
 +\chi_{\rm fin}+O(\epsilon).
 \label{chi-local-fit}
\end{equation}
The coefficients $A_i$ are local subtraction constants, while $\chi_{\rm fin}$ is the finite helicoidal susceptibility in this chosen scheme. The finite constant extracted below is therefore not claimed to be a universal number independent of the renormalization prescription. It is the finite part associated with a fixed local subtraction scheme. A different subtraction convention, for example one that keeps an explicit logarithmic local counterterm, can shift the quoted finite constant by a local scheme-dependent amount. The universal statements are instead: (i) the cancellation of the term linear in $\Omega$ for real nonchiral boundary conditions; (ii) the existence of a leading quadratic helicoidal response; and (iii) the decomposition of this response into curvature, radial-deformation, and angular--axial mixing contributions.

Odd powers of $\Omega$ may survive only if the configuration breaks the symmetry under $m\to -m$ or $k\to -k$. This could occur through chiral boundary conditions, asymmetric compactification phases, or a restricted mode sector. Such cases are not considered here.

\section{Numerical spectral benchmarks}
\label{sec:numerical}

We now provide Dirichlet numerical benchmarks and extract the finite helicoidal susceptibility using the local-subtraction prescription introduced above. The emphasis is on the physical spectral mechanisms entering the small-twist expansion: the even radial deformation, the linear angular--axial mode splitting, and the cancellation of the latter in a symmetric nonchiral vacuum sum. The finite mode sums are used only as diagnostics of symmetry and scaling, whereas the finite susceptibility is obtained only after the local power subtraction described in Eq.~\eqref{chi-local-fit}. We also include a basic finite-volume validation of the untwisted Bessel limit and test the robustness of the finite part by varying the truncation, the cutoff window, and the nonminimal coupling.

For the numerical problem we write the radial equation in Sturm--Liouville form,
\begin{equation}
 -(rF')'+r\left(\frac{m^2}{r^2}+\Omega^2k^2r^2\right)F
 =\lambda rF,
 \label{sl-form}
\end{equation}
with regularity at the origin and Dirichlet boundary condition $F(R_0)=0$. We discretize \eqref{sl-form} on a cell-centered finite-volume grid,
\begin{equation}
 r_i=\left(i+\frac12\right)h,
 \qquad
 h=\frac{R_0}{N},
 \qquad
 i=0,1,\ldots,N-1.
\end{equation}
The inner face has zero radial area, which implements the regularity condition at $r=0$, whereas the outer ghost value is set to zero to impose the Dirichlet condition. This gives a symmetric generalized eigenvalue problem,
\begin{equation}
 A\mathbf F=\lambda B\mathbf F,
 \label{generalized-numerical}
\end{equation}
where $B$ is the positive diagonal mass matrix associated with the weight $r\,dr$. In the figures below we use $R_0=1$, $L_z=2\pi$, $\mu=1$, $\xi=0$, and $N=360$, unless stated otherwise.

\begin{figure}[t]
\centering
\includegraphics[width=\columnwidth]{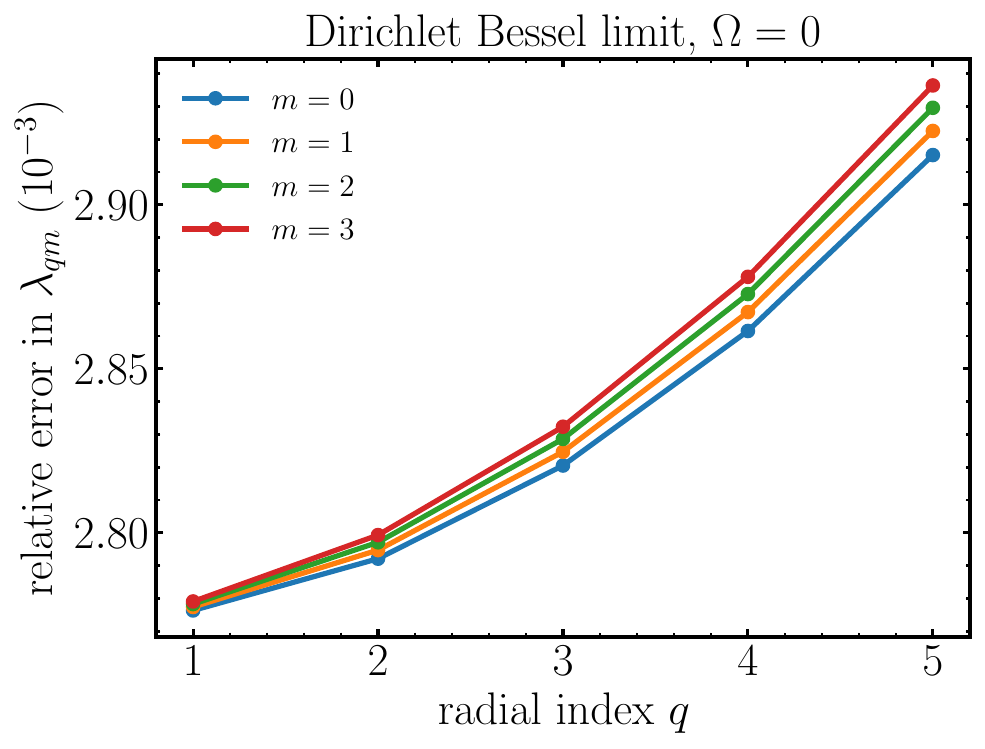}
\caption{Validation of the finite-volume radial operator in the untwisted Dirichlet limit. The plotted quantity is the relative error, in units of $10^{-3}$, between the numerical eigenvalues and the exact Bessel spectrum $\lambda_{qm}=j_{|m|,q}^2/R_0^2$.}
\label{fig:validation}
\end{figure}

Figure~\ref{fig:validation} tests the most basic requirement of the numerical scheme: in the limit $\Omega=0$ the radial operator must reduce to the ordinary cylindrical Dirichlet problem. The exact spectrum is
\begin{equation}
 \lambda_{qm}(0)=\frac{j_{|m|,q}^2}{R_0^2},
 \label{bessel-spectrum}
\end{equation}
where $j_{|m|,q}$ is the $q$-th zero of $J_{|m|}$. The figure shows that the first five radial eigenvalues in the sectors $m=0,1,2,3$ are reproduced with relative errors of order $10^{-3}$ on the grid used here. This plot is used only as a validation of the finite-volume operator and of the finite-$\Omega$ spectral benchmarks. The finite helicoidal susceptibility extracted below is computed from the analytic Bessel spectrum and the local subtraction of the corresponding mode sums, not from these finite-volume errors. The small increase of the error with $q$ is the expected finite-grid behavior: higher radial modes have shorter wavelengths and are more sensitive to the discretization.

The second benchmark isolates the radial part of the helicoidal deformation. The operator term $\Omega^2k^2r^2$ is positive and even in $\Omega$, so perturbation theory predicts
\begin{equation}
 \lambda_{qmk}(\Omega)-\lambda_{qm}(0)
 =\Omega^2k^2\langle r^2\rangle_{qm}+O(\Omega^4).
 \label{lambda-shift-numerical}
\end{equation}
For the first Dirichlet mode with $m=1$ and $k=1$, the Bessel integral gives $\langle r^2\rangle_{11}=1/3$. A finite-volume fit gives
\begin{equation}
 \lambda_{111}(\Omega)-\lambda_{11}(0)
 \simeq 0.3343\,\Omega^2-9.68\times10^{-4}\,\Omega^4,
\end{equation}
which agrees with the perturbative coefficient at the level expected from the grid resolution.

\begin{figure}[t]
\centering
\includegraphics[width=\columnwidth]{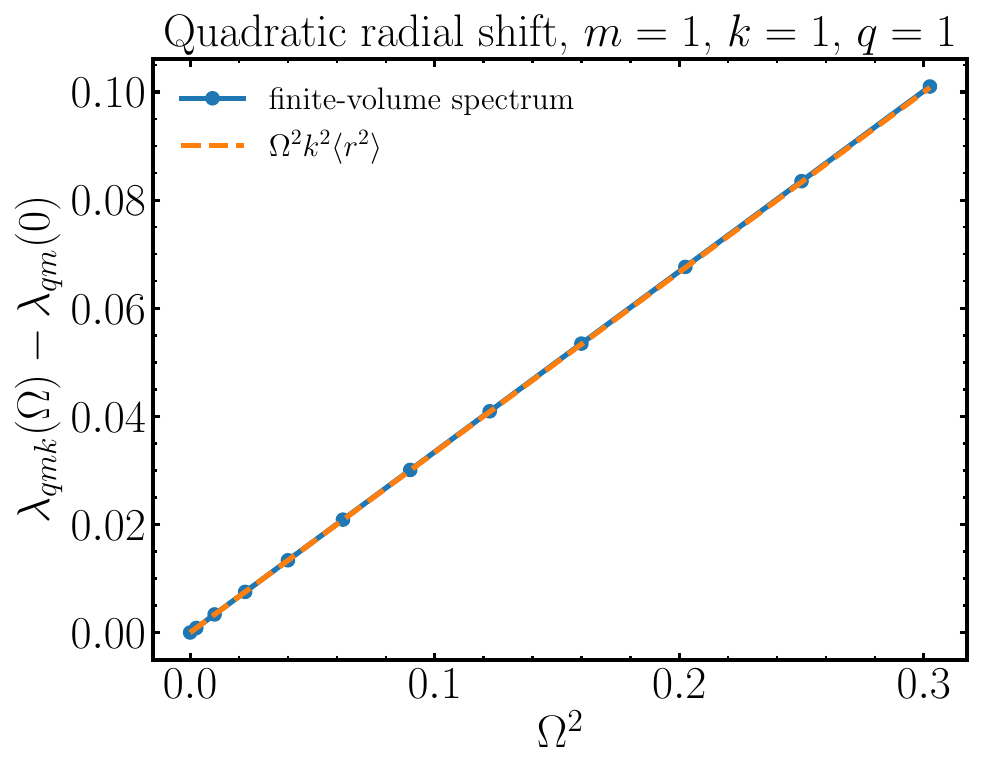}
\caption{Quadratic radial eigenvalue shift for the first Dirichlet mode with $m=1$ and $k=1$. The numerical eigenvalue shift follows the perturbative prediction $\Omega^2 k^2\langle r^2\rangle$ at small twist, confirming that the radial operator contributes an even-in-$\Omega$ correction.}
\label{fig:radial-shift}
\end{figure}

Figure~\ref{fig:radial-shift} displays this comparison. The numerical curve is almost indistinguishable from the perturbative prediction over the plotted range. This result is important because it separates the radial deformation from the chiral angular--axial splitting: the radial contribution alone cannot generate a term linear in $\Omega$ in the spectrum.

The frequency spectrum also contains the explicit angular--axial term $-2\Omega m k$. Since the radial eigenvalue depends only on $|m|$, $|k|$, and $\Omega^2$, the squared-frequency splitting between the sectors $m$ and $-m$ is exactly
\begin{equation}
 \omega_{qmk}^2-\omega_{q,-m,k}^2=-4\Omega m k.
 \label{splitting-exact}
\end{equation}
This identity is independent of the details of the radial discretization, provided the same radial eigenvalue is used for the two sectors.

\begin{figure}[t]
\centering
\includegraphics[width=\columnwidth]{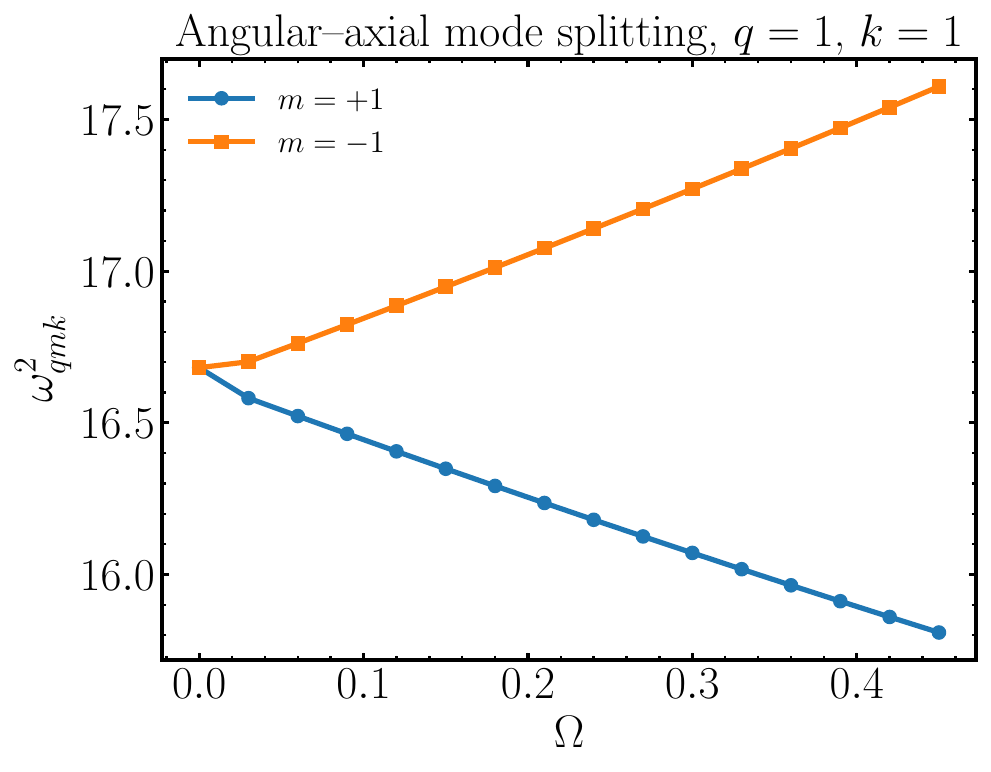}
\caption{Angular--axial splitting of the squared frequency for the first Dirichlet mode with $|m|=1$ and $k=1$. The branches with $m=+1$ and $m=-1$ move in opposite directions as the helicoidal twist is increased.}
\label{fig:splitting}
\end{figure}

Figure~\ref{fig:splitting} shows the two branches for $m=+1$ and $m=-1$. The downward branch corresponds to the sector for which $-2\Omega m k$ lowers $\omega^2$, whereas the opposite sector is shifted upward. Therefore, individual modes are sensitive to the orientation of the helicoidal twist already at first order in $\Omega$. This is the microscopic origin of the chiral splitting discussed in Sec.~\ref{sec:casimir}.

\begin{figure}[t]
\centering
\includegraphics[width=\columnwidth]{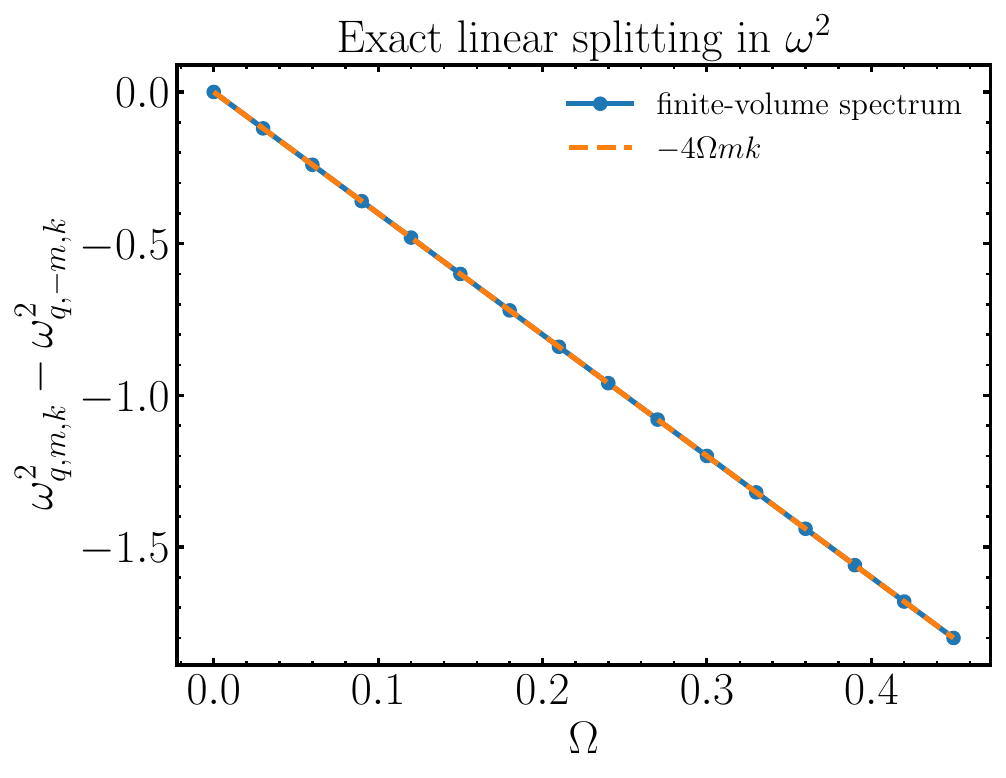}
\caption{Difference between the two squared-frequency branches shown in Fig.~\ref{fig:splitting}. The numerical data follow the exact relation $\omega_{qmk}^2-\omega_{q,-m,k}^2=-4\Omega m k$, confirming the linear angular--axial splitting.}
\label{fig:splitting-difference}
\end{figure}

Figure~\ref{fig:splitting-difference} makes the previous statement quantitative by plotting the difference between the two branches. The agreement with Eq.~\eqref{splitting-exact} is at machine precision in the numerical run. This confirms that the linear splitting is not a fitting artifact; it follows directly from the metric-induced coupling between the azimuthal and axial quantum numbers.

As a final diagnostic, we define a finite symmetric zero-point sum,
\begin{equation}
 E_{\rm trunc}(\Omega)
 =\frac12\sum_{q=1}^{q_{\max}}
 \sum_{m=-M}^{M}
 \sum_{n_z=-K}^{K}
 \omega_{qm n_z}(\Omega),
 \label{truncated-energy}
\end{equation}
with $k=n_z$ for $L_z=2\pi$. The corresponding difference
\begin{equation}
 \Delta E_{\rm trunc}(\Omega)=E_{\rm trunc}(\Omega)-E_{\rm trunc}(0)
\end{equation}
is finite by construction and is used only as a symmetry check. It should not be identified with the renormalized Casimir energy. For $q_{\max}=4$, $M=3$, and $K=3$, the numerical data in the interval $0\leq\Omega\leq0.08$ are fitted by
\begin{equation}
 \Delta E_{\rm trunc}(\Omega)\simeq 6.9033\,\Omega^2,
\end{equation}
with a maximum absolute residual of order $10^{-7}$.

\begin{figure}[t]
\centering
\includegraphics[width=\columnwidth]{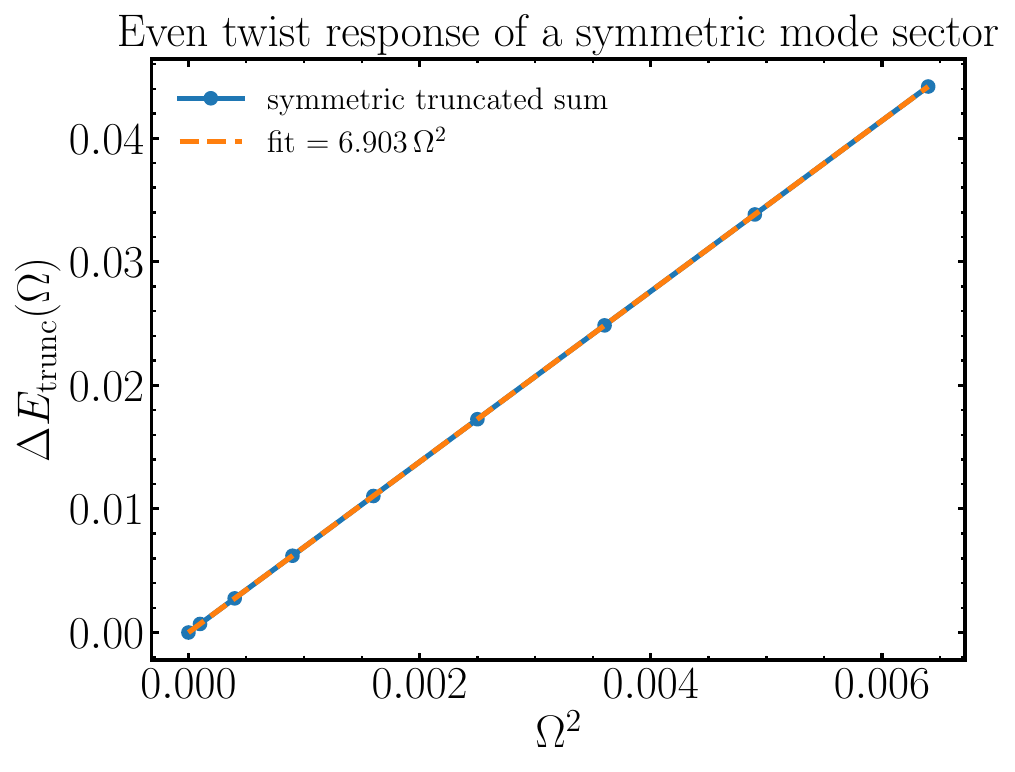}
\caption{Finite symmetric mode-sum diagnostic for the twist response. The plotted quantity is not the renormalized Casimir energy; it demonstrates that, after summing symmetrically over $m$, $-m$, $k$, and $-k$, the linear angular--axial splitting cancels and the leading response is quadratic in $\Omega$.}
\label{fig:truncated-response}
\end{figure}

Figure~\ref{fig:truncated-response} confirms the expected even response of a nonchiral mode sector and provides the numerical counterpart of Eq.~\eqref{DeltaE-quadratic}. It verifies the cancellation mechanism behind the absence of a linear term, while the renormalized energy itself still requires the analytic continuation or heat-kernel subtraction discussed in Sec.~\ref{sec:casimir}.

\begin{figure}[t]
\centering
\includegraphics[width=\columnwidth]{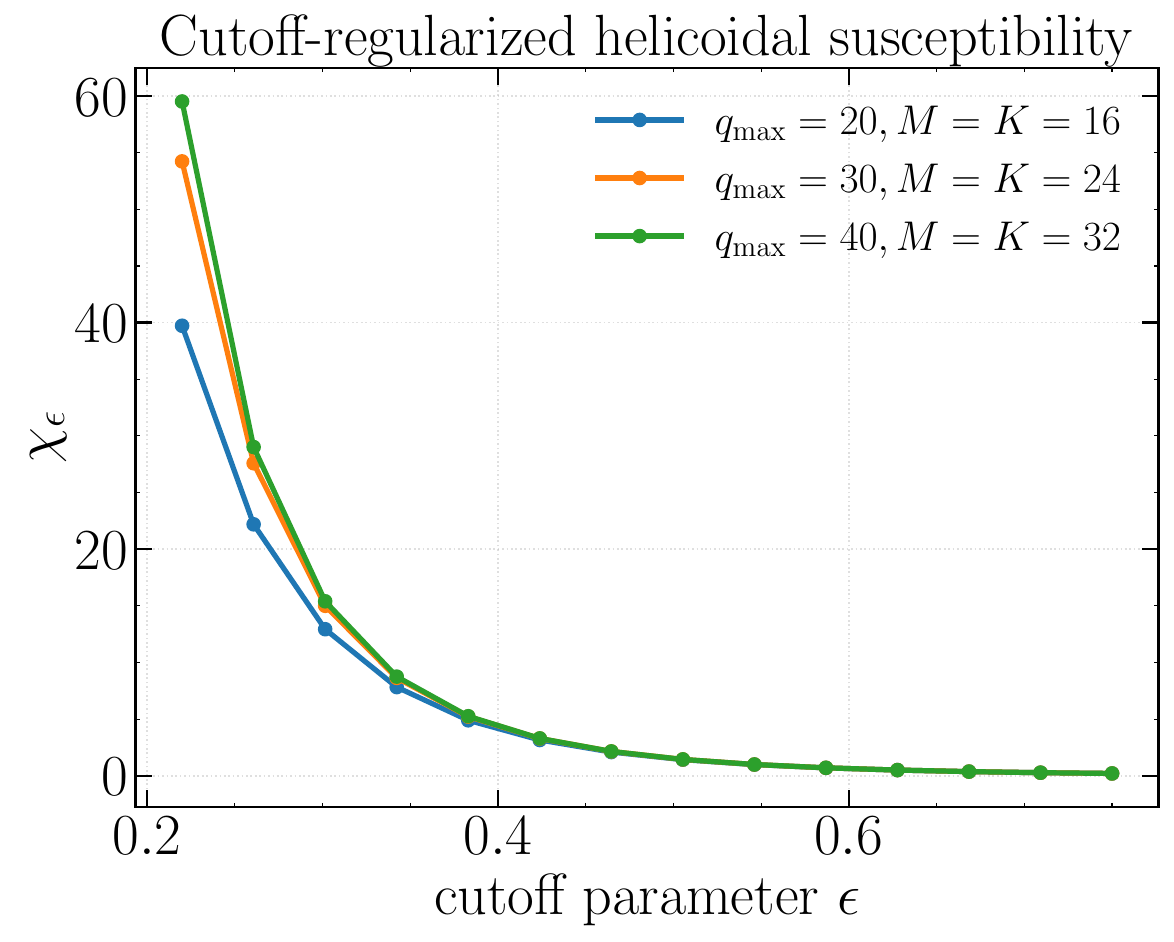}
\caption{Cutoff-regularized helicoidal susceptibility computed from the Dirichlet Bessel spectrum at $\Omega=0$. The curves show convergence with increasing symmetric truncation in the displayed cutoff window. The limit $\epsilon\to0$ requires subtraction of local heat-kernel terms and is not taken directly.}
\label{fig:cutoff-susceptibility}
\end{figure}

We finally evaluate the cutoff-regularized susceptibility \eqref{chi-cutoff} directly from the untwisted Dirichlet Bessel data. This gives a direct diagnostic of the coefficient of $\Omega^2$ in the small-twist response. Figure~\ref{fig:cutoff-susceptibility} shows $\chi_\epsilon$ for increasing symmetric truncations. The curves stabilize in the displayed cutoff window, while their growth as $\epsilon$ decreases reflects local bulk and boundary terms rather than a breakdown of the helicoidal expansion.

\begin{figure}[t]
\centering
\includegraphics[width=\columnwidth]{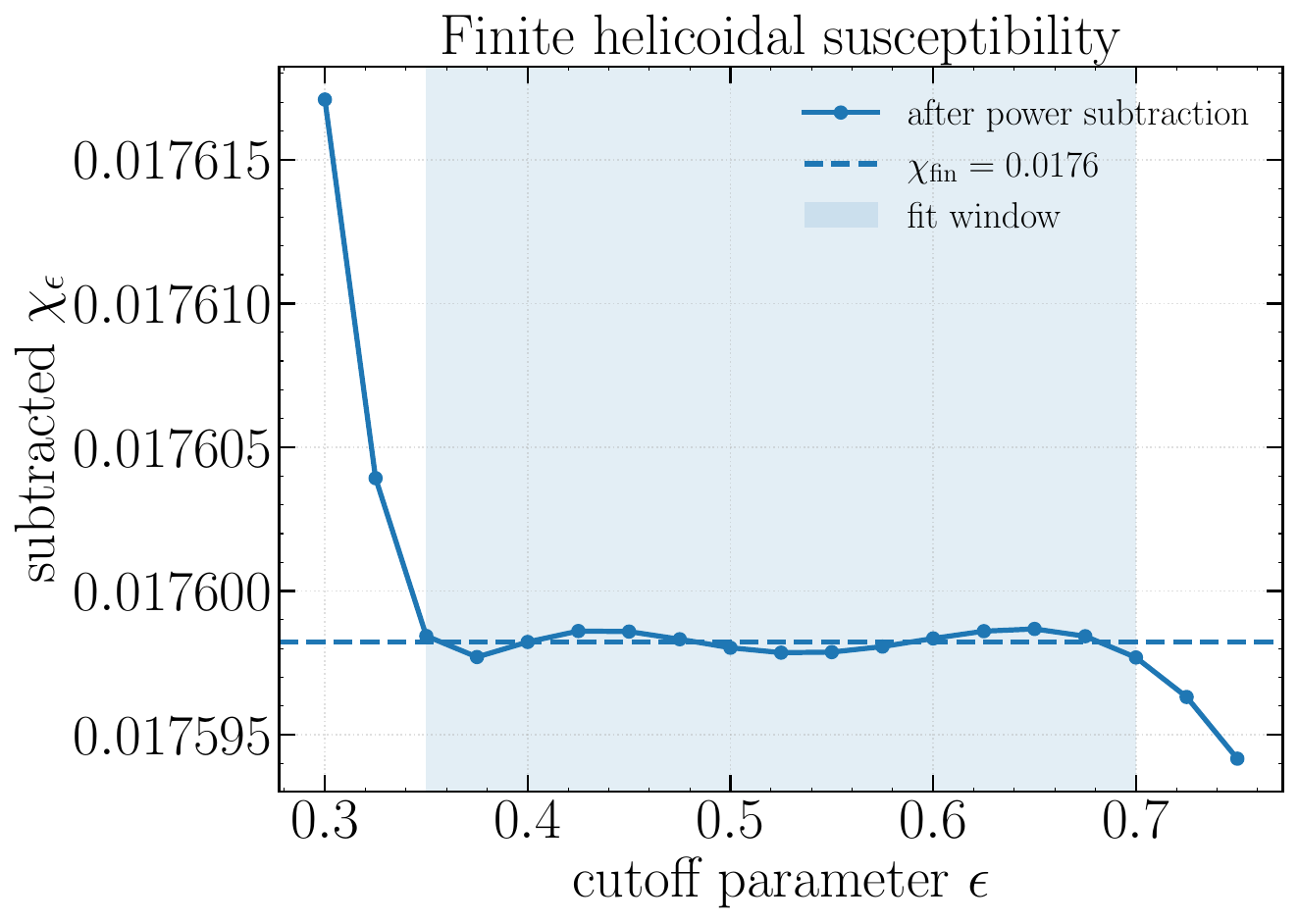}
\caption{Finite-part extraction of the helicoidal susceptibility after subtracting the local power terms in Eq.~\eqref{chi-local-fit}. The shaded region indicates the representative fit window $0.35\leq\epsilon\leq0.70$. The horizontal dashed line gives the representative value $\chi_{\rm fin}^{\rm rep}=0.01760$ for $R_0=1$, $L_z=2\pi$, $\mu=1$, and $\xi=0$; the spread under window and truncation changes is summarized in Table~\ref{tab:chi-fin}.}
\label{fig:finite-plateau}
\end{figure}

We now implement the local subtraction in Eq.~\eqref{chi-local-fit}. For the largest basis used here, $q_{\max}=80$ and $M=K=64$, a representative fit in the window $0.35\leq\epsilon\leq0.70$ gives the central value $\chi_{\rm fin}^{\rm rep}=0.01760$, with a maximum residual of order $10^{-6}$ in the fitted window. Taking the spread under the truncation and window variations in Table~\ref{tab:chi-fin} as a systematic extraction uncertainty, we quote
\begin{equation}
 \chi_{\rm fin}\simeq (1.7\pm0.2)\times10^{-2},
 \label{chi-fin-result}
\end{equation}
for the minimal-coupling Dirichlet example. The uncertainty in Eq.~\eqref{chi-fin-result} is not a statistical error bar; it quantifies the finite-window sensitivity of the local subtraction prescription. Figure~\ref{fig:finite-plateau} shows the corresponding subtracted data for the representative window. After the power-divergent part is removed, the result approaches a nearly constant finite plateau in the same cutoff range.

The stability of the extraction is summarized in Table~\ref{tab:chi-fin}. Increasing the symmetric truncation and varying the fitting window changes the finite part only mildly once the cutoff window is chosen where the mode sum is converged and the local expansion is accurate. The last digits in the table should therefore not be interpreted as universal physical precision; the robust conclusion is that, within this subtraction prescription, the minimal-coupling Dirichlet example gives a positive finite helicoidal susceptibility of order $10^{-2}$ in the dimensionless units used here.

\begin{table}[t]
\caption{Stability of the finite part extracted from Eq.~\eqref{chi-local-fit}. The representative value used in Fig.~\ref{fig:finite-plateau} corresponds to the fourth row.}
\label{tab:chi-fin}
\begin{ruledtabular}
\begin{tabular}{cccc}
$q_{\max}$ & $M=K$ & fit window & $\chi_{\rm fin}$ \\
60 & 48 & $0.35$--$0.70$ & $0.01527$ \\
70 & 56 & $0.35$--$0.70$ & $0.01739$ \\
80 & 64 & $0.32$--$0.72$ & $0.01778$ \\
80 & 64 & $0.35$--$0.70$ & $0.01760$ \\
80 & 64 & $0.38$--$0.68$ & $0.01728$ \\
\end{tabular}
\end{ruledtabular}
\end{table}

As a final robustness check we vary the nonminimal coupling. This is physically relevant because the scalar curvature is itself controlled by the helicoidal twist, $R=-2\Omega^2$, so the curvature coupling contributes directly to the quadratic response. Table~\ref{tab:xi-dependence} shows that the finite part changes approximately linearly with $\xi$ in the representative Dirichlet setup. The minimal case is positive, whereas the conformal value $\xi=1/6$ gives a negative finite part within the same subtraction scheme. Thus the sign of the helicoidal susceptibility is not a numerical artifact of the spectral discretization; it encodes the competition between radial confinement, angular--axial mixing, and curvature coupling.

\begin{table}[t]
\caption{Dependence of the finite helicoidal susceptibility on the nonminimal coupling. The extraction uses $q_{\max}=80$, $M=K=64$, $R_0=1$, $L_z=2\pi$, $\mu=1$, and the fit window $0.35\leq\epsilon\leq0.70$.}
\label{tab:xi-dependence}
\begin{ruledtabular}
\begin{tabular}{ccc}
$\xi$ & $\chi_{\rm fin}$ & max. residual \\
$0$ & $0.01760$ & $5.42\times10^{-7}$ \\
$1/12$ & $0.00358$ & $1.17\times10^{-6}$ \\
$1/6$ & $-0.01044$ & $2.86\times10^{-6}$ \\
$1/4$ & $-0.02445$ & $4.55\times10^{-6}$ \\
\end{tabular}
\end{ruledtabular}
\end{table}

\section{Casimir force and discussion}
\label{sec:discussion}

The radial Casimir force associated with the cylindrical boundary is
\begin{equation}
 F_{\rm Cas}=-\frac{\partial E_{\rm Cas}}{\partial R_0}.
 \label{force-def}
\end{equation}
At small twist,
\begin{equation}
 \Delta F_{\rm Cas}(\Omega)=
 -\Omega^2\frac{\partial\chi_{\rm Cas}}{\partial R_0}+O(\Omega^4).
\end{equation}
Thus, once a finite susceptibility is defined within a chosen subtraction scheme, its radius dependence determines the leading helicoidal correction to the radial force. This relation is useful because it converts the spectral response into a directly interpretable mechanical quantity associated with the cylindrical boundary.

\begin{figure}[t]
\centering
\includegraphics[width=\columnwidth]{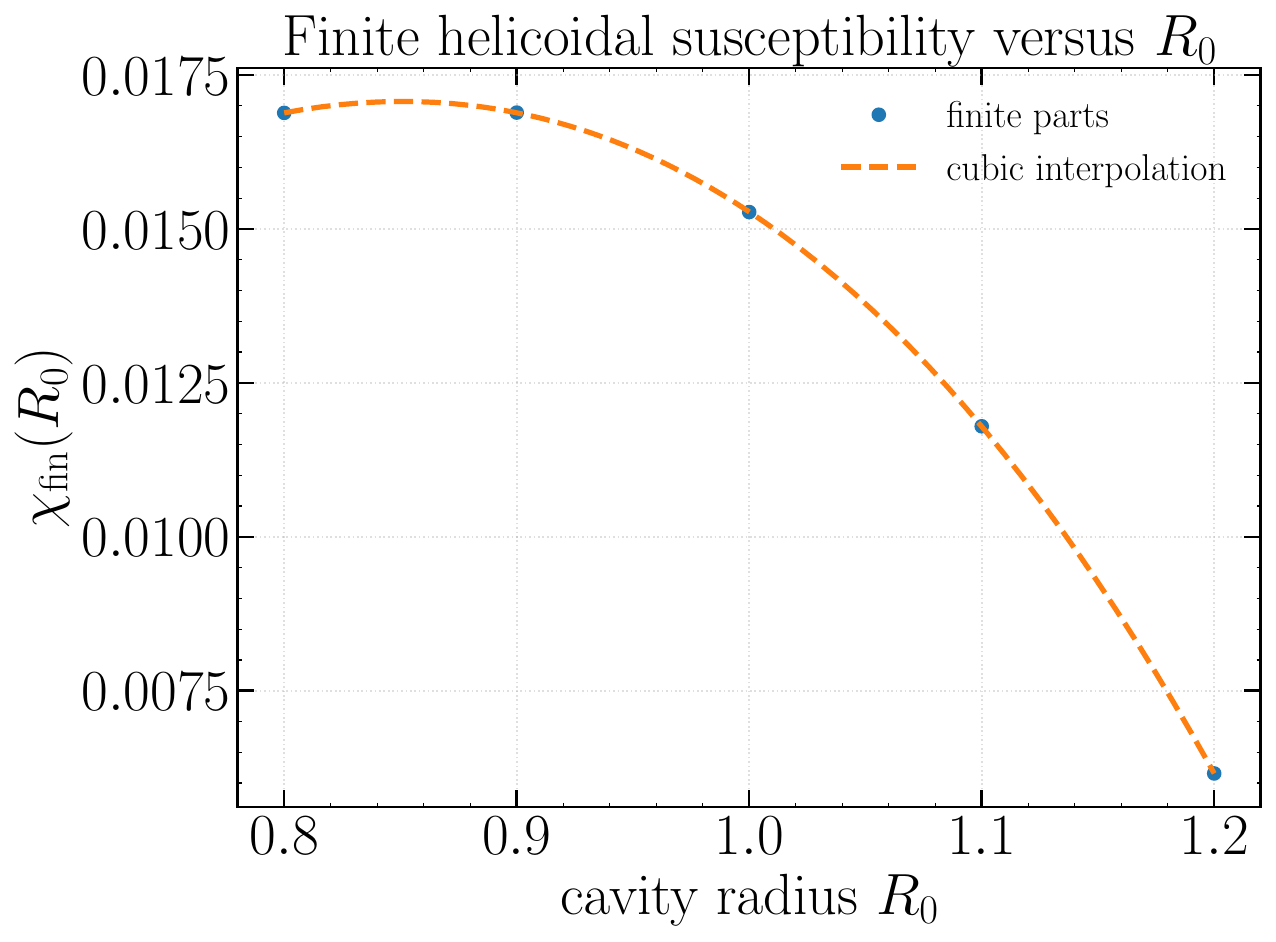}
\caption{Finite helicoidal susceptibility extracted by the local power-subtraction scheme as a function of the cavity radius. The dashed curve is a cubic interpolation used to estimate the force-response coefficient $\mathcal F_\chi=-d\chi_{\rm fin}/dR_0$.}
\label{fig:chi-r0}
\end{figure}

To estimate this correction in the Dirichlet example, we repeat the same local power-subtraction procedure for several cavity radii while keeping $L_z=2\pi$, $\mu=1$, and $\xi=0$. The result is shown in Fig.~\ref{fig:chi-r0}. A cubic interpolation of the extracted finite parts gives, at $R_0=1$,
\begin{align}
 \left.\frac{d\chi_{\rm fin}}{dR_0}\right|_{R_0=1}
 &\simeq -2.50\times10^{-2},
 \\
 \mathcal F_{\chi}(1)
 \equiv -\left.\frac{d\chi_{\rm fin}}{dR_0}\right|_{R_0=1}
 &\simeq 2.50\times10^{-2}.
 \label{force-coeff-num}
\end{align}
A central finite-difference estimate gives $\mathcal F_{\chi}(1)\simeq2.55\times10^{-2}$, providing a useful check of the interpolation. This agreement is at the level expected from the finite-$R_0$ sampling and should be regarded as a representative scheme-dependent force coefficient. Thus, within the present local-subtraction scheme,
\begin{equation}
 \Delta F_{\rm Cas}(\Omega)
 \simeq \Omega^2\mathcal F_{\chi}(1),
 \qquad R_0=1,
\end{equation}
for small twist. The positive sign means that, in the representative Dirichlet example, the helicoidal correction increases the outward force coefficient defined by $-\partial E_{\rm Cas}/\partial R_0$.

A complementary nonperturbative extension may proceed directly from the determinant \eqref{spectral-determinant}. One computes the eigenfrequencies at finite $\Omega$, forms the regularized difference with the untwisted spectrum, subtracts the local heat-kernel contributions, and differentiates with respect to $R_0$. Equivalently, the force may be obtained from the radial stress evaluated at the boundary, as in standard calculations of boundary-induced vacuum polarization.

\medskip
\noindent\emph{Physical interpretation.}---The helicoidal parameter produces a deformation distinct from that of a conical spacetime. In a cosmic-string background, the angular deficit modifies the effective angular quantum number. Here, the twist couples angular and axial modes and introduces an explicit $mk$ splitting in the spectrum. This is precisely the spectral mechanism behind chiral splittings and geometric confinement in related helicoidal quantum systems~\cite{Silva2026OQE,SilvaNettoFurtado2008}; in the present work it is promoted to a vacuum-fluctuation effect. The helicoidal cavity is therefore a useful benchmark for a type of geometric Casimir response that cannot be captured by a deficit angle alone.

This distinction is central to the physical interpretation and is supported by the numerical benchmarks. Figures~\ref{fig:radial-shift}--\ref{fig:splitting-difference} show that the helicoidal background affects the spectrum in two complementary ways: the radial eigenvalue is shifted by an even term proportional to $\Omega^2 k^2$, whereas the full frequency contains a linear angular--axial splitting proportional to $\Omega m k$. Figure~\ref{fig:truncated-response} then shows how the linear terms cancel in a symmetric nonchiral sum. Figures~\ref{fig:cutoff-susceptibility} and \ref{fig:finite-plateau} show that the corresponding susceptibility can be computed from the untwisted spectrum and that a finite part can be extracted after local power subtraction. Tables~\ref{tab:chi-fin} and \ref{tab:xi-dependence} test the stability of this finite part and display its dependence on the curvature coupling. Figure~\ref{fig:chi-r0} then connects the finite susceptibility to the force response through its radius dependence. Thus individual modes feel the helicoidal geometry linearly in $\Omega$, but the total vacuum energy of a nonchiral system is even in $\Omega$ at leading order. The quantity $\chi_{\rm Cas}$ therefore measures the quadratic response of zero-point fluctuations to a torsionless metric twist after the usual local subtractions have been implemented.

The setup also clarifies the role of the boundary. We are not considering a field inside a helicoidal material surface. The cavity is cylindrical, while the ambient geometry is helicoidal. This is precisely what makes the model useful: the boundary is simple enough to define a clean self-adjoint problem, while the ambient geometry is rich enough to generate curvature, radial deformation, and angular--axial mixing.

Several extensions are natural. One may compare Dirichlet, Neumann, and Robin conditions, introduce a compactification phase along the axial direction, or study fermionic and electromagnetic fields. Another direction is the numerical evaluation of $\Delta E_{\rm Cas}(\Omega)$ beyond the small-twist approximation, where the full determinant rather than the perturbative susceptibility would be used.

\section{Conclusions}
\label{sec:conclusion}

We have formulated the Casimir-response problem for a massive scalar field confined in a cylindrical cavity embedded in a torsionless helicoidal spacetime. The geometry is ultrastatic and cylindrically symmetric, but contains a metric twist that mixes angular and axial directions while preserving a separable self-adjoint spectral problem.

The scalar spectrum is modified by three effects: a curvature shift, an oscillator-like radial term, and an angular--axial coupling proportional to $mk$. The Dirichlet benchmarks verify the Bessel limit, the quadratic radial eigenvalue shift, and the exact linear splitting between the $m$ and $-m$ sectors. Under real self-adjoint nonchiral boundary conditions, the linear contribution in the twist cancels in the total vacuum energy. The leading robust correction is quadratic in $\Omega$ and defines a helicoidal Casimir susceptibility. We evaluated this coefficient with a smooth cutoff and extracted a finite part by subtracting the local power-divergent terms dictated by the ultraviolet structure of the mode sum. The scheme-independent content is the absence of the linear vacuum term and the quadratic structure of the response; the quoted finite constant is the value obtained in the fixed subtraction prescription used here. In the minimal-coupling Dirichlet example, this finite part is positive and of order $10^{-2}$, more precisely $(1.7\pm0.2)\times10^{-2}$ in the dimensionless units $R_0=1$, $L_z=2\pi$, and $\mu=1$. Varying the nonminimal coupling can change its sign, confirming the role of curvature coupling in the response. The radius dependence of the same scheme-defined finite part gives a positive twist-induced force coefficient at $R_0=1$.

The main lesson is that the helicoidal twist acts as a controllable geometric handle on vacuum fluctuations: it produces a linear chiral splitting at the level of individual modes, but a quadratic and mechanically measurable response at the level of the nonchiral Casimir energy. This makes helicoidal cavities a useful setting for exploring geometry-induced modifications of Casimir forces beyond purely conical or topological backgrounds, and provides a basis for future studies of spinor, electromagnetic, and finite-twist effects.

\begin{acknowledgments}
The author acknowledges support from Conselho Nacional de Desenvolvimento Cient\'{i}fico e Tecnol\'{o}gico (CNPq) (grants 306308/2022-3), Funda\c c\~ao de Amparo \`a Pesquisa e ao Desenvolvimento Cient\'{i}fico e Tecnol\'{o}gico do Maranh\~ao (FAPEMA) (grants UNIVERSAL-06395/22), and Coordena\c c\~ao de Aperfei\c coamento de Pessoal de N\'{i}vel Superior (CAPES) - Brazil (Finance Code 001).
\end{acknowledgments}

\end{document}